\documentclass[a4paper,fleqn,usenatbib]{mnras}

\usepackage{newtxtext,newtxmath}

\usepackage[T1]{fontenc}
\usepackage{ae,aecompl}

\usepackage{graphicx}	
\usepackage{amsmath}	
\usepackage{amssymb}	

\title[Magnetic instabilities in the Orion A filament]{Magnetic tension and instabilities in the Orion A integral shaped filament}

\author[Schleicher \& Stutz]{
Dominik R.G. Schleicher$^{1}$\thanks{E-mail: dschleicher@astro-udec.cl}
Amelia Stutz$^{1}$
\\
$^{1}$Departamento de Astronom\'ia, Facultad Ciencias F\'isicas y Matem\'aticas, Universidad de Concepci\'on, Av. Esteban Iturra s/n Barrio Universitario, Casilla 160-C, Concepci\'on, Chile
}

\date{Accepted XXX. Received YYY; in original form ZZZ}

\pubyear{2017}

\begin{document}
\label{firstpage}
\pagerange{\pageref{firstpage}--\pageref{lastpage}}
\maketitle

\begin{abstract}
The Orion nebula is a prime example of a massive star-forming region in our galaxy. Observations have shown that gravitational and magnetic energy are comparable in its integral shaped filament (ISF) on a scale of $\sim1$~pc, and that the population of pre-main sequence stars appears dynamically heated compared to the protostars. These results have been attributed to a slingshot mechanism resulting from the oscillation of the filament \citep{Stutz16}. In this paper, we show that radially contracting filaments naturally evolve toward a state where gravitational, magnetic, and rotational energy are comparable. While the contraction of the filament will preferentially amplify the axial component of the magnetic field, the presence of rotation leads to a helical field structure. We show how magnetic tension can give rise to a filament oscillation, and estimate a typical timescale of $0.7$~million years for  the motion of the filament to the position of maximum displacement, consistent with the characteristic timescale of the ejected stars. Furthermore, the presence of helical magnetic fields is expected to give rise to magneto-hydrodynamical instabilities. We show here that the presence of a magnetic field significantly enhances the overall instability, which operates on a characteristic scale of about $1$~pc. We expect the physics discussed here to be generally relevant in massive star forming regions, and encourage further investigations in the future.
\end{abstract}

\begin{keywords}
stars: formation -- ISM: magnetic fields -- magnetohydrodynamics
\end{keywords}



\section{Introduction}
The ''twistings and turnings of dark lanes'', which were already recognized in the photographic images of Barnard~68 \citep{Barnard05}, have since then been found and recognized in many parts of the sky, with recent successes including the Herschel Gould Belt Survey \citep{Andre10}, the Herschel Hi-GAL Milky Way survey \citep{Molinari10} and the characterization of filaments in IC~5146 \citep{Arzoumanian11}.

While many of the images were taken just recently, they should not come as any surprise, as the formation of filament is known as a natural outcome in the presence of gravity and initial perturbations \citep[e.g.][]{Zeldovich70}. It is however true that the physics governing the filaments is likely more complex than gravity alone. The presence of rotation around the filament axis has been inferred for instance for the L1641 filament in the Orion cloud \citep{Uchida91},  within Orion A  \citep{Tatematsu93, Hanawa93} and in the L1251A filament \citep{Levshakov16}. 

Polarization measurements of the magnetic field geometry probe the plane-of-the-sky projected field morphology. Due to observational restrictions, such studies have been mostly focused on the nearby low-mass clouds (but see below). Intriguingly, the plane-of-the-sky field morphology exhibits an approximately bimodal distribution of parallel and perpendicular orientations relative to the main axis of the dust filaments \citep{Li13}. Specific examples include the Serpens South cloud \citep{Sugitani11} and the B211/B213/L1495 region in Taurus \citep{Heiles00, Heyer08, Chapman11}. Optical polarization  suggests that the magnetic field is perpendicular the the B216 and B217 filaments in Taurus \citep{Moneti84, Goodman92}, while the L1506 filament is almost parallel to the direction of the magnetic field \citep{Goodman90}. These observations should ideally be complemented by Zeeman measurements to probe the magnetic field component along the line of sight, which may be necessary to confirm or discard the presence of a helical field. Signs of the presence of helical fields have been found in the dense core L1512 in Taurus \citep{Falgarone01, Hily04} and in NGC~2024 in Orion B \citep{Matthews02}. To interpret current as well as future investigations of field structures in filaments, we emphasize that detailed radiation transfer calculations including linear and circular polarization, and Zeeman line splitting will be needed to predict the observational appearance given an underlying physical structure \citep[e.g.][]{Reissl16}.

For high-mass filaments, the main work on magnetic field structures is restricted still to the Orion A cloud \citep[but see e.g.][]{Pillai15, Pillai16}. Studies of starlight polarization by \citet{Appenzeller74} have first indicated a ''magnetic pocket'' in the Orion region, which subsequently was mapped in further detail with $^{13}$CO observations by \citet{Bally91}. First signs of a helical structure were then obtained by \citet{Uchida91} and \citet{Tatematsu93} both morphologically and kinematically. Using stellar polarization, H~I Zeeman splitting and circular polarization, the helical structure was then subsequently confirmed \citep{Heiles87, Bally89, Heiles97}. {We note here that \citet{Heiles97} presented Zeeman data clearly showing the flip in the magnetic field direction on opposite sides of the filament, which is one of the most prominent observational signatures of helical fields wrapping about filaments. While in this work the author favored an interpretation of this directional flip in terms of the Eridanus shock rather than an intrinsically helical field, the interpretation in terms of helical fields was subsequently re-adopted\footnote{Press release Robishaw \& Heiles:\newline www.berkeley.edu/news/media/releases/2006/01/12$\_$helical.shtml}, though this analysis has not yet been published.} The presence of helical field structures in other high mass clouds has been indirectly inferred \citep[e.g.][]{Contreras13}, but remains to be observationally confirmed.

From a theoretical point of view, the potential relevance of magnetic fields in filaments was recognized already by \citet{Chandrasekhar53}, which has been followed up by various authors, including \citet{Nagasawa87, Fiege00, Tomisaka14} and \citet{Toci15}. Such filaments  including magnetic fields and potentially also rotation are expected to be subjected to additional magneto-hydrodynamical instabilities that go beyond gravitationally-driven fragmentation. Using a linear-stability analysis, \citet{Nakamura93} and \citet{Matsumoto94} have shown that the presence of axial or helical fields, potentially combined with the presence of rotation, may lead to the presence of a magnetically-driven mode of fragmentation \citep[see also][for the application of their models to Orion A]{Hanawa93}. 

While these calculations were initially restricted to the linear regime, \citet{Nakamura95} and \citet{Tomisaka95} subsequently tested these predictions and extended them into the non-linear regime using two-dimensional magneto-hydrodynamical simulations, thereby confirming that the results with the full set of equations. Magnetic field geometries perpendicular to the filament have been recently considered by \citet{Hanawa15}. They show that, while filaments supported by axial magnetic fields are always unstable against fragmentation due to the extremely high mass-to-flux ratios, filaments with perpendicular fields can be stable, assuming a strong enough magnetic field. \citet{Fiege00b} have confirmed the role of both gravitational and magneto-hydrodynamical instabilities, and extended them to a larger range of models. As shown by \citet{Ruediger11}, these type of magneto-hydrodynamical (MHD) instabilities may even operate within galactic disks. Most recent developments include 3D models of the filamentary structure, including turbulence and magnetic fields \citep{Seifried15}, as well as detailed predictions on their chemical structure \citep{Seifried16}. The combination of dynamical models and chemistry may indeed turn out to be relevant to probe the potential impact and strength of magnetic fields \citep{Koertgen17}.

The magnetized filament models have recently gained additional relevance, as a detailed mapping of the gravitational potential of the integral-shaped filament (ISF) in the Orion~A cloud {has} shown that magnetic and gravitational energy are comparable within $\sim1$~pc of distance from the filament \citep{Stutz16}. Such a configuration, with the observed  helical fields \citep{Heiles97}, is prone to give rise to magneto-hydrodynamical instabilities \citep{Nakamura93, Matsumoto94, Fiege00b}. In addition, both the S-shape of the ISF as well as the observed kinematics of the gas and the stars have led to the development of the ''slingshot'' paradigm, in which the velocity dispersion of the stars is enhanced through dynamical heating by an oscillating filament \citep{Stutz16}. The viability of such a process was recently demonstrated through stellar-dynamical simulations by \citet{Boekholt17}, in which the presence of an oscillating gaseous potential was found to be crucial for the heating mechanism. We note here in passing that the initial separations of the protostars, which are observed to range from several 100 to more than 1000~AU \citep{Kainulainen17}, are too large for binary interactions to play a relevant role in the heating of the velocities. Their results further suggest the presence of at least two fragmentations scales, one with periodic grouping at $50000$~AU as well as an increase of short separations below $17000$~AU. While the presence of such multi-scale fragmentation is unexpected in the context of a linear instability analysis of equilibrium filament models \citep[e.g.][]{Ostriker64}, it may be a consequence of the evolution within the non-linear regime.

Given these recent results, we aim here for an updated assessment of potential magneto-hydrodynamical effects within the ISF in the Orion A cloud. As the latter is the best-studied example for a high-mass star-forming region including high-quality data, we will here restrict ourselves to this case. It is nevertheless conceivable that similar effects occur in other high-mass star-forming regions. In general, however, these lack detailed observations of the 3D magnetic field, thus providing a central restriction to extend this type of analysis. We therefore note here that such future observations will be crucial to test if Orion A is a unique case or points toward a more general paradigm.

In section~\ref{formation} of this paper, we start with rather general considerations on the formation of a massive filament, showing that the latter naturally leads to a configuration where gravitational, magnetic and rotational energy become comparable within the central region. The latter is shown to be compatible with the data available on the ISF. We show in section~\ref{oscillation} how both magnetic tension and gravity will contribute about equally to the oscillation of such a filamentary structure, and demonstrate that the timescale of the oscillation is compatible with the dynamical timescales implied by the kinematical observations \citep{Stutz16}, as well as with the oscillation timescale inferred by \citet{Boekholt17} via stellar-dynamical calculations. An assessment of potential MHD instabilities is then given in section~\ref{instabilities}, and a discussion along with our main conclusions is presented in section~\ref{discussion}.

\section{Energetic considerations}\label{formation}

Filament formation is not yet understood, and current ideas include a variety of potential scenarios, such as colliding flows \citep{Heitsch06, Semadeni07, Banerjee09}, supersonic turbulence \citep{MacLow04, Moeckel15, Federrath16}, gravitational collapse \citep{Peters12}, or instabilities like the nonlinear thin-shell instability \citep{Vishniac94}, the thermal instability \citep{Field65} or the Kelvin-Helmholtz instability. These scenarios are also not mutually exclusive, and in reality, a mixture of different processes may give rise to the observed filament properties.

Here, we do not aim to resolve the question regarding the origin of the filaments, but aim to pursue general considerations that may apply independent of the specific scenario. We further show that the resulting conclusions are consistent with our current knowledge of the ISF.

\subsection{Gravitational and thermal energy}

We assume that a filament like the ISF in Orion forms from an initially lower-density gas along an elongated structure of size $L$, which collapses preferentially along the radial direction. Let us assume that this structure has an initial mass $M$ and a projected radius $\varpi$. As long as the collapse occurs in the radial direction, perpendicular to its axis, the line density $\Lambda=M/L$ will be constant. This is independent even of whether the filament forms due to gravitational collapse or external compression.

Adopting the mean density \begin{equation}
\bar{\rho}=\frac{M}{\varpi^2\pi L}=\frac{\Lambda}{\varpi^2 \pi} \label{Poisson}
\end{equation}
of this structure, we can estimate the gravitational potential $V$ of the filament, which is given by the Poisson equation\begin{equation}
\nabla^2 V=4\pi G \rho.
\end{equation}
As the typical length scale of spatial variations is given by the width of the filament, we can estimate $\nabla\sim\frac{1}{2\varpi}$ and approximate $\rho\sim\bar{\rho}$. Inserting these in Eq.~\ref{Poisson}, we find\begin{equation}
V\sim 16\pi G\Lambda.
\end{equation}
Considering an infinitesimal mass element $dM$ in the filament, its gravitational energy is given as $-V\,dM$. The total gravitational energy of the filament thus follows as\begin{equation}
U_G\sim- VM=-V\Lambda L=-16\pi G\Lambda^2 L,
\end{equation}
implying that $U_G/L=$const. This ratio thus depends only on the line density $\Lambda$. In particular, it is independent of the filament width $\varpi$ and the length $L$, and therefor does not increase during the collapse of the filament, as also shown by \citet{Fiege00}. This implies that the gravitational potential measured in Orion by \citet{Stutz16} had the same magnitude during earlier stages of its formation.

While the gravitational energy is thus approximately constant, we will assess in the following the behavior of other energy components. The thermal energy of the structure is given as\begin{equation}
U_T=\frac{\bar{\rho}}{\mu m_H}k_B T \varpi^2 \pi L\quad \rightarrow\quad \frac{U_T}{L}=\frac{\bar{\rho}}{\mu m_H}k_B T \varpi^2 \pi,
\end{equation}
where $\mu$ is the mean molecular weight, $m_H$ is the mass of the hydrogen atom, $k_B$ the Boltzmann constant and T the gas temperature. 

As a result of radial contraction, the mean density in the filament will increase as $\bar{\rho}\propto\varpi^{-2}$. The temperature evolution is more complex; we will first consider a case where cooling remains efficient, and the gas temperatures remains isothermal. In this case, the thermal energy will evolve as $U_T\propto \varpi^{-2}\varpi^0\varpi^2=const$, so also the amount of thermal energy will be constant within the structure. This implies that if initially the thermal energy is insufficient to balance gravity, later it will remain so.

However, if we consider a situation where the opacities are strongly increasing with density, so that the gas becomes optically thick and the effective evolution is adiabatic, we have a scaling relation of $T\propto \bar{\rho}^{2/3}$. This implies that $U_T\propto\varpi^{-2}\varpi^{-4/3}\varpi^2=\varpi^{-4/3}$. In this case, even if the thermal energy was originally subdominant, it may potentially become comparable to the gravitational energy of the filament. To test whether this may be the case, we estimate both energy components within the ISF. On a scale of 1~pc, the mean number density is about $2700$~cm$^{-3}$. Adopting a temperature of $10$~K and a length scale $L$ of 7~pc, the resulting thermal energy is about $2.2\times10^{45}$~erg. On the same scale, the gravitational potential measured by \citet{Stutz16} is about $6.3$~$({\rm km/s})^2$, thus implying a gravitational energy of about $3.4\times10^{47}$~erg. The latter gives a difference of two orders of magnitude, so that we can {safely} exclude thermal pressure as the main stabilizing agent.

For turbulent energy, one can apply relatively similar considerations. The turbulent Mach number during gravitational collapse typically adjusts itself to values comparable to the sound speed \citep[e.g.][]{Federrath11}, implying that the latter becomes comparable to the thermal energy, unless turbulence is driven by external forces. While this may initially be the case at the onset of filament formation, and potentially at late evolutionary stages due to feedback from star formation, we still expect that the turbulence will rather decay once a stable filament has formed, and therefore won't be able to support it against gravity, at least not for long.

\subsection{Magnetic fields and rotation}

As a more interesting case, we now consider the evolution of magnetic fields during filament formation. {In general, it can be expected that the magnetic field initially includes both mean and fluctuating components. We focus here predominantly on the evolution of the mean components. We denote the axial magnetic field component as $B_z$, the azimuthal component as $B_\phi$ and a potential perpendicular component as $B_\varpi$. We consider in the following the evolution of these components during the radial contraction.} The latter will amplify the magnetic field components perpendicular to the gas motions, but not the parallel ones. We can therefore expect an increase of the axial and the azimuthal field ($B_z$ and $B_\phi$), but not of the perpendicular component $B_\varpi$. We expect that at least initially, the magnetic flux will be conserved, as even low ionization degrees are typically sufficient to couple the gas and the magnetic field \citep{Pinto08, Susa15}. For the axial magnetic field, the magnetic flux is given as $B_z\varpi^2\pi$=const, thus implying that $B_z\propto\varpi^{-2}$. For the azimuthal field, the corresponding surface to consider is a section perpendicular to the field orientation, implying a magnetic flux of $B_\phi L\varpi$. In this case, we have $B_\phi\propto\varpi^{-1}$. In both cases, we thus expect a significant increase of the field strength during contraction, for the axial component even more than for the azimuthal one. 

As a result, the formation of a helical magnetic field as observed in the Orion region \citep{Heiles87} is perhaps not too surprising, even if one could expect the axial field to be considerably stronger than the azimuthal one in such a configuration. Before addressing this issue, we will now briefly estimate how the magnetic energy in both components evolves during filament contraction. The energy in either of these components is given as\begin{equation}
E_{B,\phi/z}=\frac{B_{\phi/z}^2}{8\pi}L\varpi^2\pi \quad \rightarrow \quad \frac{E_{B,\phi/z}}{L}=\frac{B_{\phi/z}^2}{8\pi}\varpi^2\pi .
\end{equation}
The energy in the axial field thus evolves as $E_{B,z}\propto \varpi^{-4}\varpi^2=\varpi^{-2}$, implying a strong increase during contraction. For the azimuthal field, on the other hand, we obtain $E_{B,\phi}\propto\varpi^{-2}\varpi^2=$const. As a result, only the axial magnetic energy is increasing during contraction, and it increases quite significantly. Unless contraction stops earlier due to other mechanisms, one will then quite  naturally reach a state where magnetic and gravitational energy become comparable. As noted by \citet{Stutz16}, this is indeed the case for the ISF in Orion on scales of about 1~pc.

Returning to the observed helical structure, a natural mechanism to produce a helical field out of an initially axial one would be the presence of rotation. Considering the angular momentum of the structure,\begin{equation}
L_{\rm ang}=M\varpi^2\Omega,
\end{equation}
with $\Omega$ the rotational angular velocity, we would expect the angular velocity of the filament to also increase as $\varpi^{-2}$ as a result of radial contraction. \citet{Tatematsu93} inferred rotation within the Orion~A cloud using the Nobeyama 45 m telescope, with $\Omega\sim1.0-1.5\times10^{-13}$~s$^{-1}$. This is consistent with the velocity gradient in the $^{13}$CO position-velocity diagram presented by \citet{Stutz16} in the direction perpendicular to the filament. At a projected radius of 1~pc, the inferred angular velocity implies a rotational velocity of about $v_{\rm rot}\sim3$~km/s. The corresponding rotational energy is thus $E_{\rm rot}\sim\frac{1}{2}I\Omega^2$, with the inertial moment $I=\frac{1}{2}M\varpi^2$. As a result, we obtain $E_{\rm rot}\sim1.2\times10^{47}$~erg, thus corresponding to about $35\%$ of the gravitational energy. The presence of such a rotational component can naturally give rise to the formation of a helical field out of an initially axial one. 

In the presence of such components, it is necessary to consider the virial equilibrium in the cloud. For this purpose, we adopt here the virial equation derived by \citet{Fiege00}, which we evaluate here assuming a homogeneous filament, and we include rotational energy as an additional term. With $V_F$ the volume of the filament, the latter gives rise to the following equation:\begin{eqnarray}
2PV_F &-& 2P_S V_F  +\frac{1}{4\pi}B_z^2 V_F\\
 &-& \frac{1}{4\pi}\left(B_{\phi, S}^2+B_{z,S}^2\right)V_F+2E_{\rm rot}+U_G=0,\nonumber
\end{eqnarray}
where $P$ denotes the pressure inside the filament, $P_S$ the external pressure at the surface, and $B_{\phi,S}$ and $B_{z,S}$ the surface components of the magnetic fields. Given that both rotational and magnetic energy contain relevant fractions of the gravitational energy, the surface terms may be relevant for the overall energy balance. We already note here that such a configuration is very likely to give rise to magneto-hydrodynamical instabilities, as we will show further below.

\section{Filament oscillations}\label{oscillation}

In this section, we explore the physical mechanism that may give rise to the filament oscillation proposed by \citet{Stutz16}, which is required to dynamically heat up the stellar population to match the observed kinematics. While the scenario has been tested using stellar-dynamical simulations \citep{Boekholt17}, we here explore the underlying cause for the oscillation of the filament.

\subsection{The magnetic tension force}
The presence of a helical field gives rise to magnetic tension once the filament is bent. In the following, we work out its contribution and the resulting timescales.
\subsubsection{Straight filament}
Before yet considering an oscillation, we start with some general considerations regarding a straight filament with a helical magnetic field. For simplicity, we will assume that the filament is aligned with the z-axis. At fixed coordinates $x$ and $y$, the magnetic field is then given as\begin{equation}
\vec{B}=\left(B_0\cos\left(\frac{z}{d_h}+\phi\right), B_0\sin\left(\frac{z}{d_h}+\phi\right), B_z\right),\label{ansatz}
\end{equation}
where $B_0$ denotes the magnetic field component perpendicular to the filament, $B_z$ the component parallel to the filament, $d_h$ is the characteristic length scale for the winding of the magnetic field in the $z$-direction, $\varpi$ as above the projected radius of the filament and $\phi$ a phase that depends on the x-y-coordinates. While the local magnetic tension at a given point will depend on the x-, y- and z-dependence of the magnetic field, we note that only the z-dependence will contribute a net force after averaging over the volume, and therefore we will only consider the z-dependence in the following, and assume $\phi=0$ without loss of generality (this can always be ensured via an appropriate choice of the coordinate system). 

Based on geometrical considerations, we also have the following relation between the magnetic field components $B_0$ and $B_z$:\begin{equation}
\frac{B_z}{B_0}=\frac{d_h}{\varpi}.\label{geom}
\end{equation}
The force density of the magnetic tension in its general form is given as\begin{equation}
\vec{f}=\frac{1}{4\pi}(\vec{B}\cdot\nabla)\vec{B}\label{tension}.
\end{equation}
Inserting (\ref{ansatz}) into (\ref{tension}), we obtain\begin{equation}
\vec{f}=\frac{1}{4\pi}B_z\partial_z \vec{B}=\frac{1}{4\pi}\frac{B_zB_0}{d_h}\left(-\sin\left(\frac{z}{d_h}\right),\cos\left(\frac{z}{d_h}\right),0  \right).
\end{equation}
The magnitude of the local force density is thus given as\begin{equation}
|\vec{f}|=\frac{1}{4\pi}\frac{B_zB_0}{d_h}=\frac{1}{4\pi}\frac{B_0^2}{\varpi},
\end{equation}
where we have used (\ref{geom}) for the second identity. As long as the filament is straight, the different contributions of the magnetic tension force will however cancel out when integrating over the volume.

\subsubsection{Bent filament}
In the next step, we consider a bent filament where the different contributions will not cancel out, but due to the bending a net force remains that aims to straighten the filament. We assume in the following a filament of length $L$, with the upper part being deflected by a length $d$. The net force density that remains after an average over the volume will thus be proportional to $d/L$, implying a net force density \begin{equation}
f_{\rm net}=\frac{1}{4\pi}\frac{B_0^2}{\varpi}\frac{d}{L}.
\end{equation}
For a filament with mass $M$ and a volume $V=\varpi^2\pi L$, the motion is then governed by the following equation:\begin{equation}
M\ddot{d}=-\frac{1}{4}\frac{B_0^2}{\varpi}\frac{d}{L}\varpi^2 L=-\frac{B_0^2 \varpi}{4}d.
\end{equation}
The latter corresponds to an oscillator equation with frequency\begin{equation}
\omega=\sqrt{\frac{B_0^2 \varpi}{4 M}}
\end{equation}
and an oscillation time\begin{equation}
T=\frac{2\pi}{\sqrt{{B_0^2 \varpi/(4 M)}}}=4\pi\sqrt{\frac{M}{B_0^2 \varpi}}.\label{time}
\end{equation}

\subsection{Application to the integral shaped filament}

{In the following, we aim to apply the previous relations to the ISF in Orion. As a first consideration, we note the S-shape of the filament, implying that it consists of two parts that are bent into different directions. The above equations are thus applied to each part separately, with each part corresponding to a length $L\sim3.5$~pc. We consider the mass per unit length at a projected radius of $1$~pc, given as $300$~M$_\odot$~pc$^{-1}$, and thus corresponding to a total mass of $1050$~M$_\odot$ for the filament \citep{Stutz16}. At a projected radius of $1$~pc, the magnetic field strength can be expected to be $\sim80$~$\mu$G \citep{Falgarone08}. 

Using equation~(\ref{time}), we obtain an oscillation period of $2.9$~million years, or about $0.7$~million years to go from a straight filament to maximum displacement. While of course this corresponds to an order-of-magnitude estimate, it is comparable to the travel time of the disk stars shown in Fig.~10 by \citet{Stutz16}. The average dynamical time of disk stars is about 0.6~Myrs, while those that have reached a maximum radius of 4~pc correspond to timescales of 4.8~Myrs. We note here that half an oscillation time can be sufficient for one turn-around of the filament. This is consistent with stellar-dynamical calculations by \citet{Boekholt17}, who have shown that one turn-around of the filament is already sufficient to explain the observed star distribution, deriving a similar timescale for the filament oscillation.}

\subsection{How to start the oscillation}
A relevant question could be how to start the oscillation of the filament. Indeed, if it is initially at rest, it will stay at rest without an external impact, and even if it formed more dynamically, one might expect that the oscillation will damp out over time. 

Assuming that the filament oscillates with a maximum displacement of about $1$~pc, our characteristic oscillation period implies a typical oscillation velocity of about $0.8$~km/s. Along with a filament mass of $\sim2100$~M$_\odot$ (within $\sim1$~pc of the filament axis) as given by \citet{Stutz16}, the linear momentum of the filament corresponds to about $1600$~M$_\odot$~km/s. Such momentum could have been induced if the filament was hit by a cloud or gas stream of about $100$~M$_\odot$, i.e. less than $10\%$ of the filaments own mass, with an average velocity of $16$~km/s. The latter corresponds to a typical velocity within the ISM. If such a collision indeed happened, it may have triggered or accelerated the star formation event.

Alternatively, one could consider that the linear momentum was not transferred within one single event, but rather results from a sequence of interactions with the interstellar medium. Assuming that each event involves a mass of $M_e=10$~M$_\odot$ with a velocity $v_e=10$~km/s, the transferred linear momentum after $N$ events would be\begin{equation}
P_e=N^{1/2}M_e v_e.
\end{equation}
As a result, one would require about $256$ events.This implies an event every $\sim4000$~years, and appears somewhat implausible. 

A more plausible explanation may involve the presence of an inhomogeneous gravitational field, which could give rise to an inhomogeneous acceleration of the filament, effectively inducing a bent structure. In such a case, both the internal magnetic fields and the self-gravity of the filament would induce an oscillation as described above. Such inhomogeneous fields may be naturally expected within the inhomogeneous medium of a galaxy, and could partly be produced by inhomogeneities within Orion itself, including for instance the Orion Nebula Cluster (ONC). While velocity gradients on $\sim3$~pc scales have recently been interpreted as signatures of infall into the cluster at the center of the ISF \citep{Hacar17}, these velocities may in fact point toward an inhomogeneous acceleration of the filament possibly caused or amplified by the gravity of the cluster, and only perhaps to second order reflect gas infall. As recently described by \citet{Stutz17}, the position of the velocity caustic is in fact offset from the cluster center by about $0.24$~pc, implying a deviation from radial infall.

\section{Magneto-hydrodynamical instabilities}\label{instabilities}
The presence of magneto-hydrodynamical instabilities in filaments with helical magnetic fields is well-established in the literature. Using a linear-stability analysis, \citet{Nakamura93, Matsumoto94} and \citet{Fiege00} have shown that in filaments with helical fields, there is not only the self-gravitational instability leading to fragmentation, but also a magnetically driven fragmentation mode. The models by \citet{Nakamura93} and \citet{Matsumoto94} not only included magnetic fields, but also the effect of rotation, and these have been applied to the Orion~A cloud by \citet{Hanawa93}. While these models initially explored the linear regime, the non-linear stage has subsequently been tested with numerical simulations, showing the non-linear fragmentation process as a result of magnetic instabilities \citep{Nakamura95, Tomisaka95}. 

Here, we return to the model developed by \citet{Nakamura93} and \citet{Matsumoto94}, taking into account the magnetic field measurements by \citet{Heiles97} as well  as the mass profile obtained by \citet{Stutz16}. Specifically, we adopt the dispersion relation of the axisymmetric instability. As previously shown, the model can be solved analytically for a pitch angle $\theta=0$, where the latter is defined asymptotically via $\tan^{-1}(B_\phi/B_z)$. The latter corresponds to the limit of a purely axial field, and leads to the dispersion relation\begin{equation}
\omega_{\rm disp}^2=-4\pi G\bar{\rho}\frac{k_z H}{1+k_zH}\frac{0.89+1.4\alpha}{1+1.25\alpha}+c_s^2k_z^2.\label{disp}
\end{equation}
We note that the latter was shown to yield reasonable agreement with numerical calculations also for other values of $\theta$ \citep{Nakamura93, Matsumoto94}. The characteristic length scale $H$ is defined as\begin{equation}
H^2=\frac{c_s^2}{4\pi G\bar{\rho}}\left( 1+\frac{1+\cos^2\theta}{2}\alpha+\beta \right),\label{scale}
\end{equation}
with the parameters
\begin{equation}
\alpha=\frac{B^2}{8\pi\bar{\rho}c_s^2}
\end{equation}
and
\begin{equation}
\beta=\frac{2\Omega^2H^2}{c_s^2}.
\end{equation}
The projected radius $\varpi$ corresponds to  $2\sqrt{2}$ times the characteristic length $H$, implying thus $H=0.35$~pc for $\varpi=1$~pc. We adopt here a sound speed of $0.26$~km/s within the filament, corresponding to a temperature of $10$~K, an average number density of $6.7\times10^3$~cm$^{-3}$ and a magnetic field strength of $80$~$\mu$G. With these parameters, we find a value of $\alpha\sim16.8$. The $\beta$ parameter describes the influence of rotation, and we adopt here the angular velocity of $\sim10^{-13}$~s$^{-1}$ measured by \citet{Tatematsu93}, implying $\beta\sim32.6$.

As a result, we can determine the wavenumber of maximum instabililty, which is given as \citep{Hanawa93}\begin{equation}
k_{\rm z,max}=0.72H^{-1}[(1+\alpha+\beta)^{1/3}-0.6].
\end{equation}
With the parameters given above, we find $k_{\rm z,max}=6.4$~pc$^{-1}$, corresponding to a fragmentation scale of $2\pi/k_{\rm z,max}\sim0.98$~pc due to magneto-hydrodynamical instabilities. The characteristic mass of the fragment is then about $300$~M$_\odot$, leading to the formation of a small stellar cluster. This may have provided for instance an initial seed for the ONC, which subsequently may have grown through further accretion from the filament.

\begin{figure}[htbp]
\begin{center}
\includegraphics[scale=0.7]{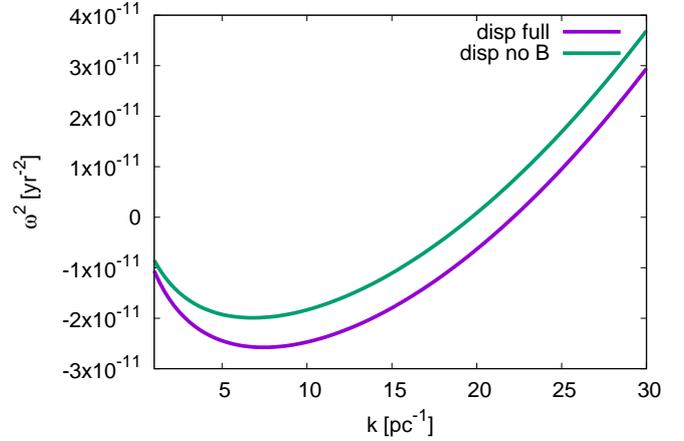}
\caption{The dispersion relation for fragmentation in the integral-shaped filament of Orion, based on Eq.~\ref{disp}. We compare the dispersion relation including the current measurements of the magnetic field and the expected dispersion relation in their absence. Due to the presence of the helical field, the instability is somewhat enhanced, and shifted toward larger wavevectors.}
\label{figdisp}
\end{center}
\end{figure}

We compare the full dispersion relation to the expected dispersion relation without a magnetic field in Fig.~\ref{figdisp}. From the dispersion relation, we infer at typical timescale of $1.4$~million years for the instability, and it is clearly visible that the instability is stronger in the presence of a helical field, as the dispersion relation is systematically shifted to more negative values. The latter also implies that the instability extends both to somewhat smaller and larger scales as a result of magneto-hydrodynamical processes. In particular, the smallest scale on which the filament is still unstable corresponds to about $0.3$~pc, similar to the larger of the two characteristic fragmentation scales inferred by \citet{Kainulainen17}. This may suggest that the larger of the two fragmentation scales may be produced by filament instabilities, while the smaller fragmentation scales can be due to the non-linear evolution. As an additional cautionary remark, which was already pointed out by \citet{Kainulainen17}, one should note that the filament models considered in this work assume an equilibrium structure of the filament, while in fact the true filament may not be in an equilibrium state. This non-equilibrium state implies that to fully capture the fragmentation properties of the observed system will require future work utilizing non-linear simulations.

\section{Discussion and conclusions}\label{discussion}
In this paper, we have explored the potential relevance of magnetic tension and instabilities in the Orion A integral-shaped filament (ISF). The motivation for this work has been the detailed observations of this system, both in terms of the detailed helical field structure \citep{Heiles97} as well as the mapping of the gravitational potential and the kinematics of the gas and the stars \citep{Stutz16}. On characteristic scales of about $1$~pc, this analysis has shown that gravitational and magnetic energy are essentially comparable, thus providing a strong case for the occurrence of magneto-hydrodynamical instabilities during the evolution of the filament. 

The observed kinematics of the stars around the filament, and especially their comparison with the kinematics of protostars and the gas of the filament itself, provides strong evidence for the need to dynamically enhance the velocity dispersion of the stars. \citet{Stutz16} already suggested that an oscillating filament may provide such a heating mechanism, a hypothesis that was subsequently confirmed via stellar-dynamical simulations by \citet{Boekholt17}. Observations on the initial separation of the protostars on the other hand have shown that these are too large for binary interactions to become relevant \citep{Kainulainen17}. In a different context, focusing on the exploration of low-mass filaments, \citet{Gritschneder17} have demonstrated that such an oscillation mechanism can be obtained as a result of gravitational interactions themselves.

Here, we have started with basic considerations of energy and angular momentum conservation during the formation of filaments, showing that the latter naturally leads to a state where magnetic and rotational energy are important compared to the gravitational energy, consistent both with the above-mentioned results by \citet{Stutz16} and the measurement of the angular velocity of the filament by \citet{Tatematsu93}. We therefore expect that such a situation can be achieved independently of the specific formation mechanism of the filament. In addition, we note that signs of rotation have been found also in other star-forming filaments, including L1641 \citep{Uchida91} and L1251A \citep{Levshakov16}, and are also known from other types of filaments, such as the so-called elephant trunks in HII regions \citep[e.g.][]{Gahm06}. A 'double helix' nebula, potentially indicative of a magnetic torsional wave, has been observed by \citet{Morris06} at a distance of $100$~pc from the Galaxy's dynamical center.

We also assessed {how magnetic tension can give rise to the observed oscillation of the filament, finding an approximate timescale of $2.9$~million years in the ISF. The turnaround time of the filament is thus expected to be about $\sim1.45$~million years, consistant with the observed kinematical timescales}, as well as the stellar-dynamical calculations by \citet{Boekholt17}.

Finally, we have assessed the impact of the magnetic field on the overall instability in the filament based on the models of \citet{Nakamura93} and \citet{Matsumoto94}, showing that the dispersion relation is systematically shifted toward more negative values, and enhancing the instability already present in the linear regime. The latter may potentially explain the larger of the two fragmentation scales inferred by \citet{Kainulainen17}, though clearly an assessment of the impact of deviations from a dynamical equilibrium as well as the implications of a fully non-linear treatments will be necessary in the future. First efforts into investigating such 3D effects have already been carried out \citep[e.g.][]{Seifried15}. However, the extension to high line-masses, helical field geometries, and realistic values of the magnetic field strength are required to assess the physics of the ISF and of other massive filaments. 

To explore whether  the ISF is a special case or in fact characteristic and typical for high-mass filaments, it is essential to obtain Zeeman data of comparable quality to the \citet{Heiles97} investigation for a larger number of high-mass filaments, as only these provide simultaneously both the line-of-sight field strength and direction. When combined with polarization data probing the plane-of-the-sky field component, together these allow for the inference of the full 3D magnetic field configuration.  At the same time, important indirect probes can be obtained through detailed measurements of the kinematics of the gas and the stars. We expect that both future observations with ALMA\footnote{ALMA website: http://www.almaobservatory.org/} as well as the upcoming Gaia\footnote{Gaia website: http://sci.esa.int/gaia/} data releases will be crucial to investigate these effects in Milky Way clouds.

\section*{Acknowledgements}

We thank Tjarda Boekholt, Mike Fellhauer, Valentina Isabel Gonzalez Lobos, Andrew Gould and Rafeel Riaz for valuable discussions on the topic. AS and DRGS are thankful for funding from the ''Concurso Proyectos Internacionales de Investigaci\'on, Convocatoria 2015'' (project code PII20150171) and the BASAL Centro de Astrof\'isica y Tecnolog\'ias Afines (CATA) PFB-06/2007. DRGS acknowledges funding through Fondecyt regular (project code 1161247) and ALMA-Conicyt (project code 31160001).




\input{slingshot_v2.bbl}





\bsp	
\label{lastpage}
\end{document}